# Extraction of Relevant Images for Boilerplate Removal in Web Browsers

Joy Bose
Senior Member, IEEE
Hyderabad, India
joy.bose@ieee.org

## ABSTRACT
Boilerplate refers to unwanted and repeated parts of a webpage (such as ads or table of contents) that distracts the user from reading the core content of the webpage, such as a news article. Accurate detection and removal of boilerplate content from a webpage can enable the users to have a clutter free view of the webpage or news article. This can be useful in features like reader mode extraction or webpage summarization in web browsers. Current implementations of reader mode content extraction in major web browsers perform reasonably well for textual boilerplate content in static webpages, but are mostly based on heuristics and hence may not be correct when the webpage content is dynamic. Also, they often do not perform well for removing boilerplate content in the form of images and multimedia in webpages. For detection of boilerplate images, one needs to have knowledge of the actual layout of the images in the webpage, which is only possible when the webpage is actually rendered. In this paper we discuss some of the issues in relevant image extraction, present the design of a testing framework to measure accuracy and a classifier to extract relevant images by leveraging a headless browser solution that gives the rendering information for images.

## 1. INTRODUCTION
The problem of extracting the relevant content after removing noise from webpages is called by various names such as boilerplate removal [1-5] and content extraction. This can be useful in features of web browsers such as webpage translation, webpage summarization and reader mode. Reader mode refers to the feature in web browsers where webpages are shown in an easy to read format by removing ads and clutter. It is a classification problem, where one must classify each HTML element in the webpage as relevant, i.e. part of the main article, or not relevant.

The current implementations of the content extraction algorithm in read mode or reader mode on modern browsers such as Chrome, Firefox, Safari and Edge use heuristics [6-8] and perform reasonably well for textual content in news articles. However, the heuristics-based approaches, which involve multiple rules to decide whether an HTML is relevant or not, are not flexible enough and an old heuristic rule may become obsolete if the HTML standards and webpage layouts change. Also, most news articles are heavy in multimedia including images, videos and embedded content such as twitter feeds. The problem with classifying images is often that they are hidden inside nested div elements or iframes, which makes it difficult to extract and classify the content to decide whether it is relevant or not to display in reading view.

In this paper we make the following contributions: first we describe the design of a labeling tool and testing framework for testing the accuracy of a dataset of webpages in reading view. Secondly, we leverage a solution that uses the rendering result of a headless browser to provide features related to the DOM structure of the rendered webpage. Finally, we present the accuracy results for a curated dataset of high-impact news articles webpages.

## 2. RELATED WORK
The general problem in webpage content extraction and boilerplate removal is to extract data from a webpage in a structured way. HTML is unstructured and has many kinds of flexibility, which make extraction of meaningful information from the HTML (and the DOM tree created by the rendering from the HTML) alone quite difficult. Some of the cases where it is difficult to infer the layout from HTML alone are as follows: sometimes an HTML table may be used for layouting the webpage, at other times the DOM elements may be used to create a table layout. Getting the layout from the DOM tree of the webpage alone is therefore difficult.

Even the render tree representation of the webpage, formed by the CSS along with the HTML cannot give enough accurate information about how the final webpage would look when actually rendered in a variety of user agents and resolutions, such as a desktop browser or a mobile browser.

Various approaches [1-5] have been proposed to solve the problem of content extraction including heuristics, machine learning based approaches for classification such as decision trees, support vector machines (SVM), conditional random fields (CRF), and approaches based on image processing on the webpage layouts, such as VIPS [6].

Kohlschutter [1] found that shallow text features such as average word length, sentence length, text density and link density can lead to a pretty good guess of which part of the webpage is boilerplate or not, in standardized datasets such as CleanEval. The algorithm was adapted in an open source library called boilerpipe that inspired a few commercial implementations. In a related paper [2], Kohlschutter used a text density related feature to perform segmentation of a webpage.

Body Text Extraction or BTE is another approach, which works on the principle that blocks of relevant content would be contiguous and therefore the main challenge is to find the largest such block in a given webpage. The method is called maximum subsequence segemtation [3, 4]. A variation of this principle, splitting the HTML content by tags and using some rules to cluster the tags, has been used in the jusText [5] algorithm. Another variant is the Goldminer algorithm [6].

Weninger [7] in their WWW paper used the text to tag ratio of the lines in the webpage as a feature to classify the content as boilerplate or not, also getting good results.

Vogels [8] used deep learning using a convolutional neural network to classify boilerplate in webpages using a range of block level features, getting good results over benchmarks.

VIPS [9] is a different kind of algorithm, which instead of extracting features from the HTML tries to simulate the visual layout of the webpage, by segmenting the webpage into non-overlapping rectangular visual blocks using some heuristics. However, it is not implemented for content extraction and boilerplate removal in real time.

Most HTML pages (at least the pages of news articles or blogs) do have a kind of structure, with the relevant article content in the middle as a block, and the table of contents on the sides. The structure is common in the sense that even if we see a webpage in a language we do not understand such as Chinese or Arabic, we have no difficulty in visually understanding the different parts of the webpage or which part is boilerplate and which is the relevant content. A commercial solution diffbot [10] has leveraged this property and created an image classifier with deep learning, with an API to get the different sections of an URL such as header and footer. Search engines such as Google or Bing would also have to index some, though not all, sections of crawled webpages.

Various commercial browsers have used variations of the above mentioned approaches [11-14]. Implementations of libraries like beautifulsoup in python [15] can also perform some web scraping.

However, as in the case of diffbot and vips, a visual based approach is more likely to give good results on a wide variety of webpages. In this paper, we used a headless renderer to give us the visual rendering information, from which we extracted features to train a machine learning model.

## 3. PROBLEM STATEMENT AND APPROACHES

Our general problem is to segment the webpage into blocks or units in such a way so that each individual unit can be identified and categorized. Here the categories can be something like "boilerplate" vs "not boilerplate" or something more granular such as "heading", "article content", "table of contents" etc.

The unit of the webpage can be any one of the following: DOM node or render tree node or visual block. The first challenge is to define the unit at a level that balances accuracy and complexity. Then we have to segment the webpage into such blocks. Finally, we classify each of the units into the categories as needed.

One way to solve the problem is to take a computer vision approach to segmentation, similar to VIPS [9] and diffbot [10]. This involves getting the rendered webpage image or screenshot, starting with the pixels and segmenting the image into regions or blocks on the basis of some measure of pixel similarity. For example, one can use a clustering algorithm to cluster similar pixels. But simultaneously one also needs to have a cost function to control the number of units such that the number of regions also does not grow too much. In one top down approach, the whole webpage is treated as one region and then the non-coherent regions are divided in each step. In another approach, each pixel is treated one region and then the coherent regions are combined in each step. We need an optimal number of regions. One big challenge in this approach is how to link the block unit back to the original DOM elements, so that it can be presented to the reading view to be shown to the user.

Another way is to get some measure such as text density (number of words per line, as in [1]) instead of a purely visual measure and use it to grow the regions or identify the similar regions.

One can also perform some smoothening of the page and define (or fit a function for) some kind of Markov random field or other type of field that covers the whole webpage. Here the idea is that the blocks seem to cluster together on the webpage, so labels for neighboring blocks would be a good approximation of the label of the block under consideration.

Finally, another solution is to have a headless browser to render the webpage and produce a feature vector comprising position and other features of each HTML element. This is the approach we have taken in this paper, since it solves some of the problems with other approaches and balances complexity and accuracy.

## 4. TESTING FRAMEWORK

In order to improve the accuracy of the algorithm for content extraction and boilerplate removal in webpages, we first have to build the platform and tools to evaluate the accuracy and compare different algorithm results. So we first built a tagging or labeling tool to label the individual elements of the webpage, as well as a testing framework for webpages using Selenium. We used a dataset of around 1000 high impact URLs, incorporating factors such as frequency of usage from telemetry data, variation in webpage design etc. We decided to concentrate on news webpages initially since their usage is usually high as per Alexa top sites. Also, news articles are better structured and so are a good starting point to target accuracy improvements.

Fig.1 gives a screenshot of the tagging tool, and fig. 2 shows the architecture of the testing framework. The steps of the testing are described in the following subsections:

### 4.1 Building a tagging/ labeling tool

We built a tagging / labeling tool to tag individual elements of the webpages as relevant or not relevant. The tool was written in Javascript, implemented as a web browser extension, and generated an overlay on the DIV element, including text and image elements, on which the mouse hovered, the user then tagged the element as relevant or not. It then saved the results in JSON format, as well as the original and tagged webpage.

Fig. 1 gives a screenshot of the tagging tool.

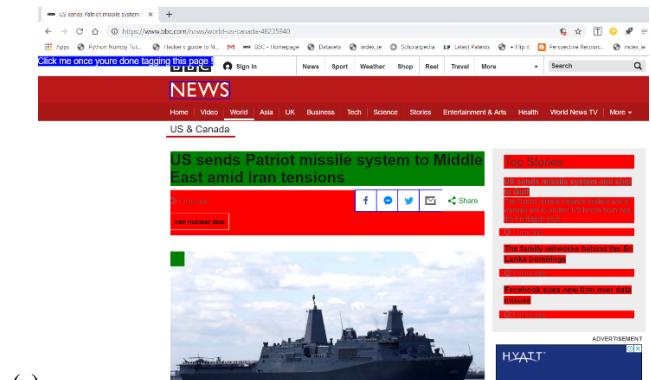

(a)

```
{
  "label":"R",
  "tag":"H1",
  "id":"240",
  "content":"<h1 class=\"story-body__h1 relevant\" id=\"240\">US sends Patriot missile"
},
{
  "label":"NR",
  "tag":"DIV",
  "id":"243",
  "content":"<div class=\"story-body__mini-info-list-and-share-row noise\" id=\"243\">"
},
```

(b)

**Figure 1. (a)Tagging tool, a browser extension to tag the news article elements as relevant (R) in green or non-relevant (NR) in red.(b)Sample output showing tagged elements.**

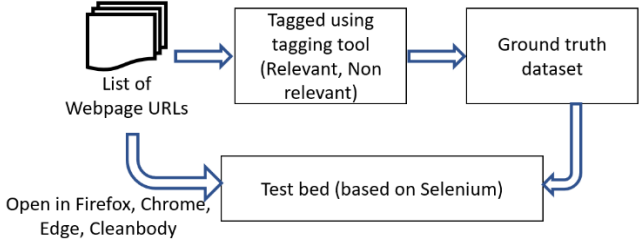

**Figure 2. Architecture of the testing framework.**

### 4.2 Generating a dataset of tagged HTML pages

Using the tagging tool, we manually tagged a list of around 1000 high relevance URLs for news articles. For each text and image element, we tagged it as relevant or not relevant. This was saved in a JSON file. The original HTML pages corresponding to the URLs were also saved.

### 4.3 Using a test bed to get accuracy

We built a test bed to measure and compare the accuracy of the reading view implementation in different browsers, including our custom algorithm. The testing framework used Selenium for automating the opening of a set of saved webpages in reading view and supported several browsers including Chrome, Firefox, and Edge.

The original framework was only meant for text. We modified the testing framework to give the performance results for images, and using the headless browser solution.

Fig. 2 gives the architecture of the testing framework

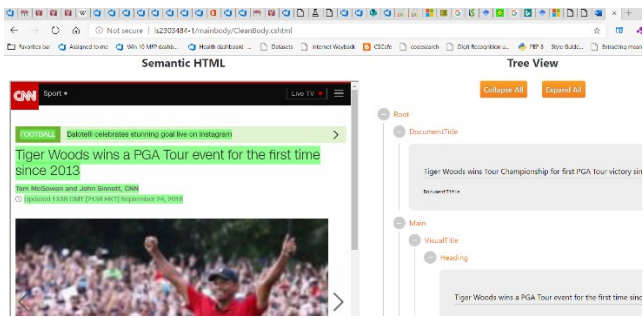

**Figure 3. Predictions generated by the headless browser tool for a given URL. Green indicates relevant content.**

## 5. EXTRACTION USING HEADLESS BROWSER SOLUTION

We chose a custom headless browser solution since it is customizable and could be easily modified to classify images.

Since this solution uses a headless browser, it can extract features related to the layout of webpage elements from the rendering engine. solution worked by rendering the webpage and generating a vector of features of each webpage element such as position, X and Y coordinates, metadata related features, relative position as a ratio and other features.

Fig. 3 shows the web interface for the solution.

We trained a custom image classifier using the solution. The classifier first used the headless browser solution to extract features such as position and size of the images, from the rendered webpage. Our classifier used an implementation of the MART gradient boosting algorithm [16, 17], called fasttree, for the classification.

Gradient boosting often performs better than other methods of shallow learning for simple classification, so we used this algorithm.

The default parameters we used for the gradient boosting are learning rate=0.4, shrinkage=0.53, number of leaves=92, iterations count=50.

We used a training:testing ratio of 70:30 in our webpage dataset for the solution.

## 6. RESULTS

The results of the accuracy of the extraction algorithms for text and images using the testbed are given in table 1.

**Table 1. Classification accuracy on different browsers for extracting relevant content from webpages**

| Browser | Precision | Recall | F1 Score |
|---|---|---|---|
| Chrome (text) | 0.91 | 0.88 | 0.88 |
| Firefox (text) | 0.91 | 0.87 | 0.87 |
| Edge (text) | 0.94 | 0.85 | 0.88 |
| Headless browser DOM solution (text) | 0.68 | 0.82 | 0.72 |
| Chrome (images) | 0.31 | 0.5 | 0.34 |
| Firefox (images) | 0.6 | 0.65 | 0.61 |
| Edge (images) | 0.94 | 0.76 | 0.79 |
| Headless browser DOM solution (images) | 0.5 | 0.6 | 0.55 |

As we can see, the read mode solutions in major browsers performs well for text, although the image results are variable. Our headless browser solution for images performs well for a few chosen dynamic webpages but poorly overall. It is still work in progress and its accuracy is expected to improve with more training and tweaking of parameters.

## 7. CONCLUSION AND FUTURE WORK

We have described a framework for testing the accuracy of extraction algorithms for read mode in web browsers. The current results show the content extraction methods in major browsers performing well for text but variably for images.

We have trained an image classifier using the headless browser enabled solution and are in the process of further improving its accuracy by tweaking the model parameters and improving the feature selection.

In future we hope to switch to the solution using a web service once the accuracy is improved sufficiently. We also intend to expand the test bed dataset to include a wider selection of webpages.

## 8. ACKNOWLEDGEMENTS
The author would like to thank Pratish Kumar and Chandana NT for valuable help in building the tagging tool and test bed, and Manoj B for guidance and advice.

## 9. REFERENCES

[1] Kohlschütter C, Fankhauser P, Nejdl W. 2010. Boilerplate detection using shallow text features. In *Proceedings of the third ACM international conference on Web search and data mining* WSDM '10, ACM.

[2] Kohlschütter, Christian, and Wolfgang Nejdl. A densitometric approach to web page segmentation. *Proceedings of the 17th ACM conference on Information and knowledge management*. ACM, 2008.

[3] Pasternack, Jeff, and Dan Roth. Extracting article text from the web with maximum subsequence segmentation. *Proceedings of the 18th international conference on World wide web*. ACM, 2009.

[4] Pomikálek, J., 2011. *Removing boilerplate and duplicate content from web corpora*. Doctoral dissertation, Masarykova univerzita, Fakulta informatiky.

[5] justext - Algorithm.wiki. Google code archive [Online]. Available: https://code.google.com/archive/p/justext/wikis/Algorithm.wiki

[6] Endrédy I, Novák A. More Effective Boilerplate Removal - the GoldMiner Algorithm. *Polibits*. 2013 Dec(48):79-83.

[7] Weninger, Tim, William H. Hsu, and Jiawei Han. CETR: content extraction via tag ratios. *Proceedings of the 19th international conference on World wide web*. ACM, 2010.

[8] Vogels, T., Ganea, O.E. and Eickhoff, C., 2018. Web2Text: *Deep Structured Boilerplate Removal.* arXiv preprint arXiv:1801.02607.

[9] Cai, D., Yu, S., Wen, J. R., & Ma, W. Y. 2003. *VIPS: a vision-based page segmentation algorithm.* Technical report MSR-TR-2003-79, Microsoft Research

[10] Diffbot: Automatic APIs [Online]. Available: https://www.diffbot.com/products/automatic/

[11] Github. Mozilla/readability. [Online]. Available: https://github.com/mozilla/readability

[12] Github. Chromium/dom-distiller. https://github.com/chromium/dom-distiller

[13] Microsoft Edge team. Reading view. 2017. DOI: https://docs.microsoft.com/en-us/microsoft-edge/dev-guide/browser-features/reading-view

[14] Daniel Aleksandersen. Web Reading Mode: Determining the main page content. [Online]. Available: https://www.ctrl.blog/entry/browser-reading-mode-content.html

[15] Geeks for Geeks. Reading selected webpage content using Python Web Scraping. [Online]. Available: https://www.geeksforgeeks.org/reading-selected-webpage-content-using-python-web-scraping/

[16] Multiple Additive Regression Trees [Online]. Available: statweb.stanford.edu/~jhf/MART.html

[17] Chris J.C. Burges. From RankNet to LambdaRank to LambdaMART: An Overview. Microsoft Technical Report MSR-TR-2010-82. June 2010